\documentclass[letterpaper,10pt,twocolumn]{article}

\usepackage[utf8]{inputenc}
\usepackage{amsmath}
\usepackage{amsfonts}
\usepackage{amssymb}
\usepackage{amsthm}
\usepackage[spanish]{babel}
\usepackage{fontenc}
\usepackage{graphicx}
\usepackage[dvips]{hyperref}

\author{J.L. Silva-Acosta\footnote{Estudiante, Unidad Académica de Física, Universidad Autónoma de Zacatecas.}, 
R. Juárez-Maldonado\footnote{Asesor, Unidad Académica de Física, Universidad Autónoma de Zacatecas.}}
\title{Propiedades Termodinámicas de Poblaciones Microbianas sobre Membranas Biológicas}
\date \today

\begin{document}
\maketitle
\begin{abstract}
En este trabajo se desarrolla un formalismo mecánico estadístico general para estudiar sistemas
restringidos a superficies de revolución, los cuales son un buen modelo para el estudio de las propiedades
termodinámicas de sistemas microbiologicos y/o macromoléculas viviendo sobre membranas biológicas como
la pared celular. 
\end{abstract}

\section{Introducción}
La gran mayoría de los fenómenos que ocurren en la naturaleza se manifiestan bajo la influencia de algún
mecanismo que los restringe a vivir en un espacio limitado. Tal es el caso por ejemplo de los organelos
celulares los cuales se encuentran confinados en el interior de la membrana celular o las diferentes
bio-moléculas que se encuentran restringidas a moverse sobre la pared celular [1,2]. Otro ejemplo es toda
clase de moléculas, seres microscópicos o partículas suspendidas en interfaces [3] (agua-aire, aceite-agua,
burbujas, etc.). La dinámica misma de la vida está restringida a la superficie del planeta, así como los
huracanes y corrientes de aire, entre muchos ejemplos más.
\linebreak 
\linebreak 
Resulta pues de fundamental importancia investigar las propiedades termodinámicas y los efectos que pueda tener
por ejemplo la curvatura sobre las propiedades dinámicas y estáticas de dichos sistemas. Se han hecho algunos
intentos por estudiar las propiedades termodinámicas de sistemas restringidos a superficies esféricas [4] y 
cilíndricas [5], también se han hecho estudios de como la curvatura afecta las propiedades de auto-difusión [6].
El formalismo de la Mecánica Estadística es lo suficientemente general para incluir tales restricciones y
efectos, sin embargo existen determinadas sutilezas a la hora de los cálculos explícitos de propiedades tales como
la presión o como las funciones de distribución radial, sobre todo cuando se incluyen los efectos de tamaño finito. 
Consecuentemente, un desarrollo detallado de las ecuaciones fundamentales de la termodinámica, es de gran importancia
para todo aquel investigador no especializado en el campo (biólogos, químicos, microbiologos, entre otros).
\linebreak 
\linebreak 
El propósito del presente trabajo es desarrollar explícitamente la estructura metodológica formal de la Mecánica
Estadística Clásica para el caso particular de sistemas restringidos a vivir sobre superficies de revolución y
aplicar este formalismo en primera aproximación a sistemas constituidos por biomoléculas; por ejemplo proteínas
sobre la membrana celular. Una versión más general que incluye el caso de sistemas multicomponente sobre
variedades Riemannianas de dimensión arbitraria está siendo preparada para su publicación [7]. 
\linebreak 
\linebreak 
El trabajo completo se organiza de la siguiente forma. En la sección 2 se plantea el formalismo canónico 
para la dinámica de sistemas restringidos a superficies de revolución. En la sección 3 se revisan los 
conceptos de Mecánica Estadística bajo la teoría de los ensambles: Canónico y Gran-Canónico. 
En esta misma sección se analiza el caso particular de gases ideales y modelos con esferas duras, 
además de derivar la fórmula barométrica. 
Finalmente en la sección 4 se da un sumario de los principales resultados del presente trabajo. 

\section{Formalismo canónico sobre superficies de revolución}
Considere una partícula de masa $m$ restringida a moverse sobre la superficie de revolución definida por
$S\left(u,\phi\right)=\left(f(u)\cos\phi,f(u)\sin\phi,g(u)\right)$ [8] y bajo la influencia del campo externo
$V=V(u,\phi)$. La dinámica de dicha partícula está completamente contenida en las ecuaciones de Hamilton [9]

\begin{equation}
 \dot{p}_{u}=-\frac{\partial H}{\partial u}
\end{equation} 
\begin{equation}
 \dot{p}_{\phi}=-\frac{\partial H}{\partial \phi}
\end{equation} 
\begin{equation}
 \dot{u}=\frac{\partial H}{\partial p_{u}}
\end{equation} 
\begin{equation}
\dot{\phi}=\frac{\partial H}{\partial p_{\phi}}
\end{equation} 

Donde $H$ es el Hamiltoniano

\begin{equation}
 H=\frac{p^{2}_{u}}{2m\left[\left(\frac{df}{du}\right)^2+\left(\frac{dg}{du}\right)^2\right]}+\frac{p^{2}_{\phi}}{2m\left[f(u)\right]^2}+V(u,\phi).
\end{equation} 

Por ejemplo, un caso de interés para la biología por su analogía con la célula, es la esfera de radio 
$R$, $f(u)=R\sin{u}$ y $g(u)=R\cos{u}$. El
Hamiltoniano es simplemente

\begin{equation}
 H=\frac{p^{2}_{u}}{2mR^2}+\frac{p^{2}_{\phi}}{2mR^{2}\sin^{2}{u}}\\+V(u,\phi).
\end{equation} 
Otro ejemplo de importancia es el toro, ya que los globulos rojos (células de la sangre) pueden ser
ser modelados en muy buena aproximación como un toro. En este caso, $f(u)=R+r\sin{u}$ y $g(u)=r\cos{u}$,
donde $R$ es el radio de giro y $r$ el radio interior. El Hamiltoniano es de la forma
\begin{equation}
 H=\frac{p^{2}_{u}}{2mr^2}+\frac{p^{2}_{\phi}}{2m[R+r\sin{u}]^2}\\+V(u,\phi).
\end{equation}

\section{Ensambles Estadísticos}
Consideremos un sistema de $N$ partículas de masa $m$ restringido a la superficie de revolución
$S\left(u,\phi \right)=\left(f(u)\cos{\phi},f(u)\sin{\phi},g(u)\right)$. Si denotamos un punto del espacio fase de este sistema por
$\left(\textbf{q},\textbf{p}\right)=\left(q_{1},…q_{N},p_{1},…p_{N}\right)=\left(u_{1},\phi_{1},…u_{N},\phi_{N},p_{u_{1}},p_{\phi_{1}},…p_{u_{N}},p_{\phi_{N}}\right)$,
el potencial de interacción entre pares de partículas con $u_{ij}=u\left(q_{i}-q_{j}\right)$ y un campo externo con
$V_{i}=V(q_{i})$, entonces el Hamiltoniano del sistema está dado por
\begin{equation}
\begin{array}{ccc}
 H(q,p)&=& \frac{1}{2m} \sum_{i=1}^{N}\frac{p^{2}_{u_{i}}}{\left[\left(\frac{df}{du}\right)^2+\left(\frac{dg}{du}\right)^2\right]} +\frac{p^{2}_{\phi_{i}}}{\left[f(u)\right]^2} \\
 &&+U_N(u,\phi)+V_N(u,\phi)
\end{array}
\end{equation}

Donde $U_N(u,\phi)=\sum^{N}_{i<j}U_{ij}$ y $V_N(u,\phi)=\sum^{N}_{i=1}V_{i}$.

A partir de este sistema podemos modelar en primera aproximación, varios tipos de sistemas biológicos, por ejemplo
un eritrocito contaminado con algún tipo de virus que reside en su superficie, como se ve en la figura 1.
\begin{figure}[!h]
\begin{center}
\includegraphics[height=6cm]{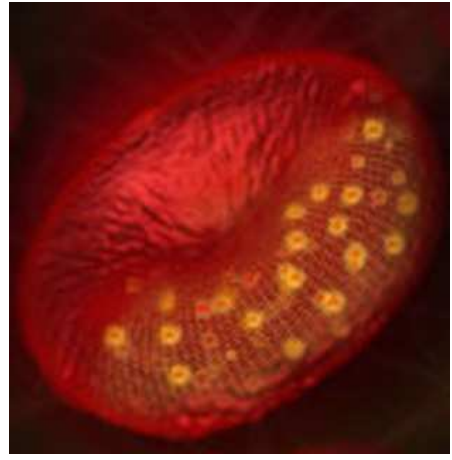}
\caption{Eritrocito (globulo rojo) contaminado con el virus de la malaria [11].}
\end{center}
\end{figure}
Este sistema se puede aproximar con la superficie de revolución del toroide y al virus con pequeñas biomoléculas
puntuales, como se ve en la figura 2.
\begin{figure}[!h]
\begin{center}
\includegraphics[height=6cm]{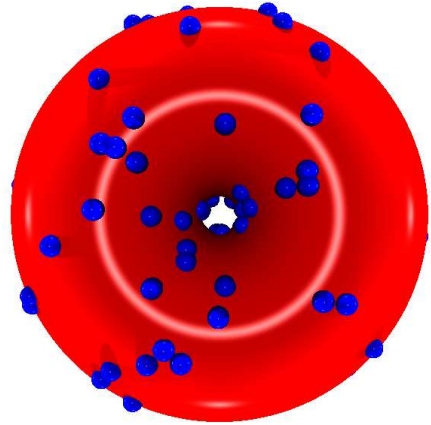}
\caption{Analogía de un eritrocito contaminado por un virus que reside en su superficie.}
\end{center}
\end{figure}
Otro ejemplo son algunos tipos de proteínas que están restringidas a vivir sobre la membrana celular. Por ejemplo 
las proteínas periféricas, de las cuales la mayoría quedan ancladas a la membrana celular por interaccines 
eléctricas (fuerzas de van der Waals) o por unión covalente de la proteína con lípidos de membrana. En la fugura 3
se muestra un ejemplo de membrana celular, la cual podemos aproximar al igual que muchos otros ejemplos más.
\begin{figure}[!h]
\begin{center}
\includegraphics[height=6cm]{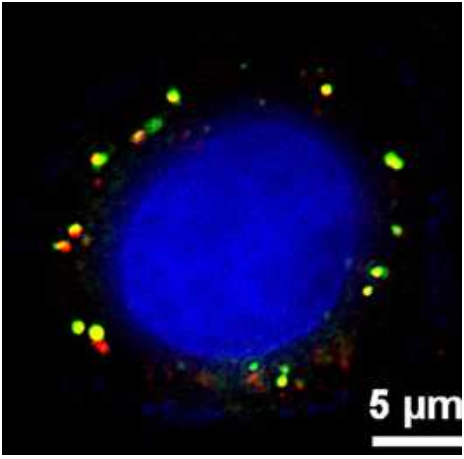}
\caption{Célula con distintos tipos de biomoléculas residiendo sobre su superficie [12].}
\end{center}
\end{figure}
Nuevamente este sistema, puede modelarse con una primera aproximación, la cual puede ser la superficie de una
esfera con biomoléculas puntuales sobre dicha superficie como se muestra en la figura 4.
\begin{figure}[!h]
\begin{center}
\includegraphics[height=6cm]{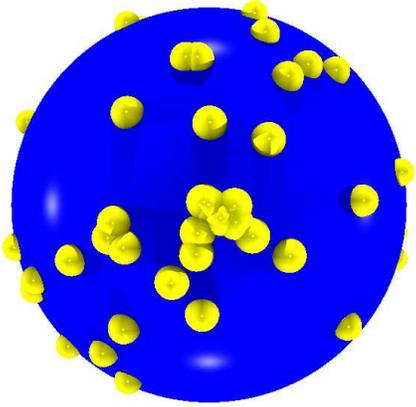}
\caption{Modelo eférico de la célula de la Figura 3.}
\end{center}
\end{figure}
\subsection{Ensamble Canónico}
En este ensamble el sistema intercambia energía térmica con el medio ambiente en el cual se encuentra inserto,
pero no materia. Es decir, está en equilibrio con un reservorio térmico a temperatura $T$, con el cual puede
intercambiar calor pero no trabajo y su número de partículas es constante. Las variables características son el
número de partículas $N$, el volumen $V$ (en estos casos es el área $A$ en consideración) y la temperatura $T$. 
Dado que el sistema no está aislado, su energía no es
constante y fluctúa (puede intercambiar energía con el reservorio), entonces, sólo podemos hablar de la
probabili\-dad de que el sistema adopte una energía determinada para un dado valor de la temperatura del
reservorio. La probabilidad buscada es

\begin{equation}
 P_{E}=\frac{e^{-\beta{E}}}{Z_{N}} 
\end{equation} 
Donde la función de partición canónica $Z_{N}$ es una constante de normalización impuesta para que la suma de las
probabilidades de todos los estados sea uno. Se define para estos sistemas en consideración como

\begin{equation}
Z_{N}=\frac{1}{N!h^{2N}}\int d^{N}qd^{N}pe^{-\beta{H}}
\end{equation} 
Siendo $H$ el Hamiltoniano del sistema, $N$ el número de partículas, $\beta=\frac{1}{kT}$ y $k$,$h$ las constantes de Boltzmann
y Plank respectivamente. Esta ecuación nos da la energía libre de Helmholtz $F=-kT\ln{Z_{N}}$ del sistema en
función de sus variables naturales, lo que supone un conocimiento termodinámico exhaustivo del sistema mediante
las ecuaciones de estado.
\begin{equation}
 p=-\left(\frac{\partial F}{\partial A}\right)_{N,T}
\end{equation} 
\begin{equation}
 S=-\left(\frac{\partial F}{\partial T}\right)_{N,A}
\end{equation} 
\begin{equation}
 U=-\left(\frac{\partial \ln{Z_{N}}}{\partial \beta}\right)
\end{equation} 
Por tanto conocer la función de partición es resolver el problema estadístico. Trabajaremos entonces con esta
función de partición hasta llevarla a una forma convencionalmente simple. 
Consideremos el sistema de N partículas que se describe al inicio de la sección III. Tomando el Hamiltoniano 
para este sistema y después de varios cálculos realizados que no se muestran para ahorrar espacio, la función
de partición canónica toma la forma
\begin{equation}
 Z_{N}=\frac{1}{N!\Lambda^{2N}}\int\prod_{i=1}^{N}dA_{i}e^{-\beta(U_N(u,\phi)+V_N(u,\phi))}
\end{equation} 
Donde $dA_{i}=\sqrt{\left[\left(\frac{df}{du_{i}}\right)^{2}+\left(\frac{dg}{du_{i}}\right)^{2}\right]\left[f(u_{i})\right]^{2}}du_{i}d\phi_{i}$ 
y $\Lambda=\frac{h}{\sqrt{2\pi{mkT}}}$ es la longitud de onda térmica de de Broglie. Esta última
expresión para la función de partición canónica es lo bastante general y simple para evaluar algunos
casos idealizados sobre superficies de interés. Además, apartir de esta expresión para la función de partición
junto a la ecuación de la presión (11) y despreciando el potencial externo $V_N=0$, se puede derivar la ecuación de estado del virial
\begin{equation}
 \frac{p}{kT}=\rho (1+B_2(A,T)\rho+\dots),
\end{equation}
la cual es una ecuación que incluye las desviaciones de la idealidad. Aquí, $\rho=N/A$ es la densidad
de moléculas por unidad de área y $B_2(A,T)$ es el segundo coeficiente del virial, este último está
dado por
\begin{equation}
 B_2(A,T)=B^{*}_2(A)-A \frac{\partial B_2^{*}(A)}{\partial A},
 \end{equation}
 donde 
 \begin{equation}
  B_2^{*}(A)=-\frac{1}{2A}\int_A\int_A dA_1 dA_2 \left[ e^{-U(q_1,q_2)/kT}-1 \right].
 \end{equation}

\subsubsection{Esfera}

Retomando la función de partición canónica y asumiendo que las biomoléculas son de un gas ideal en 
ausencia de un campo externo, entonces $U(u,\phi)=V(u,\phi)=0$ por tanto la función de partición es simplemente
\begin{equation}
 Z_{N}=\frac{A^{N}}{N!\Lambda^{2N}}
\end{equation} 
donde $A=\int{dA}=\int\sqrt{\left[\left(\frac{df}{du}\right)^2+\left(\frac{dg}{du}\right)^2\right]\left[f(u)\right]^{2}}dud\phi$
es el área de la superficie de revolución. Recordando que para ésta superficie de revolución $f(u)=R\cos{u}$
y $g(u)=R\sin{u}$ entonces el área $A=4\pi{R^{2}}$.
\linebreak  
\linebreak 
De aquí es trivial obtener las ecuaciones termodinámicas de estado a partir de la energía libre de Helmholtz. 
En particular, la ecuación de la presión, la entropía y la energía interna son 
\begin{equation}
  p=-\left(\frac{\partial F}{\partial A}\right)_{N,T}=\frac{NkT}{A}
\end{equation} 
\begin{equation}
 S=-\left(\frac{\partial F}{\partial T}\right)_{N,A}=k\ln{\frac{A^{N}}{N!\Lambda^{2N}}}+2N
\end{equation} 
\begin{equation}
 U=-\left(\frac{\partial \ln{Z_{N}}}{\partial \beta}\right)=NkT
\end{equation} 
Interesantemente, la ecuación de estado para la presión es análoga a la ecuación para un gas ideal contenido
en un volumen $V$, pero en este caso estamos hablando de un gas ideal restringido a moverse sobre una superficie.
También es notable que la energía interna solo depende de la temperatura y claro del número de particulas.
Para el toroide tenemos las mismas ecuaciones y por tanto propiedades similares pero con un área $A$ distinta
ya que al resolver la integral de superficie de revolución tenemos que $A=4\pi^{2}Rr$.
\\
El caso más sencillo no ideal, es cuando se considera un potencial de interacción a pares tipo
esfera dura
\begin{eqnarray}
 U_{ij}=U(\alpha_{ij})&=&0\, si\, \alpha < \alpha_{ij} < 2\pi - \alpha \\
 &=&\infty\, si\, 2\pi-\alpha < \alpha_{ij} < \alpha
\end{eqnarray}
donde $\alpha_{ij}$ es la separación angular entre dos partícluas cualesquiera y
$\alpha$ es el ángulo que cubre una partícula sobre la superficie esférica. Entonces la ecuación de estado del virial
queda de la siguiente forma
\begin{equation}
 \frac{p}{kT}=\rho (1+\frac{A}{8}{\alpha}\sin{\alpha}\rho+\dots)
\end{equation}

\subsection{Ensamble Grancanónico}
En este ensamble el sistema intercambia materia y energía con el medio ambiente en el cual se encuentra inserto. 
El medio ambiente actúa entonces a la vez como reservorio de temperatura y de partículas. Las variables 
características son el potencial químico $\mu$, el volumen $V$  y la temperatura $T$; Vamos a suponer que el volumen 
del sistema se mantiene constante, de manera que pensamos que el sistema se encuentra separado del reservorio 
por paredes rígidas pero permeables. Si el sistema se encuentra en equilibrio con el reservorio, el número de 
partículas no será mas constante, pero podemos asumir que tanto su energía media como su densidad media de 
partículas, permanecen constantes, de esta forma los estados accesibles del sistema son en este caso los 
autoestados de la energía para una partícula, para dos partículas, etc..
\linebreak 
\linebreak 
Al igual que en el ensamble canónico, a partir de la gran función de partición se pueden calcular expresiones 
para los valores esperados o promedios de las funciones de estado y se define para estos sistemas en 
consideración como
\begin{equation}
 \Theta(\mu,A,T)=\sum_{N=0}^{\infty}\lambda^{N}Z_{N}(A,T)
\end{equation} 
Donde $\lambda=e^{\beta\mu}$. Esta ecuación nos lleva al gran potencial que esta dado por
$\Theta=-pA=-kT\ln\Theta$, entonces es fácil demostrar que el número promedio de partículas es
\begin{equation}
 \bar{N}=kT\left( \frac{\partial \ln\Theta}{\partial \mu}\right)_{A,T}
\end{equation} 
y la energía promedio
\begin{equation}
 \bar{E}=-\frac{\partial \ln{\Theta}}{\partial \beta}+\frac{\mu}{\beta}\frac{\partial \ln{\Theta}}{\partial \mu}.
\end{equation} 
\subsection{Fórmula barométrica}
Hay que notar que la función de partición canónica para un gas ideal es
\begin{equation}
 Z_{N}=\frac{1}{N!}Z^{N}_{1}
\end{equation} 
donde 
\begin{equation}
 Z_{1}=\frac{1}{\Lambda^{2}}\int{dAe^{-{\beta}V(u,\phi)}}
\end{equation} 
es la función de partición para una sola partícula bajo la influencia
de un campo externo. En ausencia de un campo externo, la función de partición canónica del sistema es
simplemente

\begin{equation}
 Z_{N}=\frac{A^N}{N!\Lambda^{2N}}=\frac{\left(2{\pi}mkT\right) ^{N}}{N!h^{2N}}A^{N}
\end{equation} 
Como ya se había visto. Entonces, de aquí la gran función de partición es
\begin{equation}
 \Theta(\mu,A,T)=\sum^{\infty}_{N=1}\frac{\left(\frac{\lambda{A}}{\Lambda^{2}}\right)^{N}}{N!}=e^{\frac{\lambda{A}}{\Lambda^{2}}}
\end{equation} 
Por tanto, tenemos el número promedio de partículas como
\begin{equation}
 \bar{N}=kT\left(\frac{\partial \ln{\Theta}}{\partial \mu}\right)_{A,T}=\frac{\lambda{A}}{\Lambda^{2}} 
\end{equation} 
Retomando el caso del sistema bajo la influencia de un campo externo, si denotamos la densidad de partículas
por $\rho(u,\phi)$ entonces el número promedio de partículas en la superficie es

\begin{equation}
 \bar{N}=\int{dA\rho(u,\phi)}
\end{equation} 
Por otra parte, la gran función de partición es 
\begin{equation}
 \Theta(\mu,A,T)=\sum^{\infty}_{N=1}\frac{\left(\lambda{Z_{1}}\right)^{N}}{N!}=e^{\lambda{Z_{1}}}
\end{equation} 
entonces
\begin{equation}
 \bar{N}=\lambda\left(\frac{\partial \ln{\Theta}}{\partial \lambda}\right)=\lambda{Z_{1}}
\end{equation} 
Y por tanto, igualando las ecuaciones (33) y (35) y teniendo en cuenta que $Z_{1}$ está dada por la eciación (29),
obtenemos la formula barométrica.
\begin{equation}
 \rho(u,\phi)=\rho_{0}e^{-{\beta}V(u,\phi)}
\end{equation} 
donde $\rho_{0}=\frac{2\pi{mkT}}{h^{2}}e^{\beta{\mu}}$.
Esta fórmula nos dice como se distribuyen las partículas sobre la superficie bajo la influencia de un potencial
externo, lo cual puede resultar útil para determinar la respuesta de las poblaciones biológicas al ser perturbadas.

\section{Conclusiones}

Hemos derivado expresiones explicitas para la ecuación de estado de sistemas de partículas restringidas
a una superficie de revolución. En particular, obtuvimos la ecuación de estado del virial para el caso
de sistemas en contacto con un reservorio de energía (ensable canónico). Y específicamente calculamos la
ecuación de estado para un una población biológica que vive en la superficie de una célula, la cuál fue modelada
con un sistema de partículas tipo esfera dura restringido a vivir sobre la superficie de una esfera simulando
la pared celular. Además encontramos una expresión general para determinar la respuesta del sistema
ante la presencia de perturbaciones externas.


\end{document}